# Size Reduction of The Transfer Matrix of Two-Dimensional Ising and Potts Models

M. Ghaemi[1, 2] and G. A. Parsafar[3]

1- Atomic Energy Organization of Iran, Deputy in Nuclear Fuel Production,
. Tehran, IRAN
2-Chemistry Department, Teacher Training University, Tehran, Iran
Email: ghaemi@saba.tmu.ac.ir

3-College of Chemistry, Sharif University of Technology, Tehran, Iran

Email: parsafar @ sharif.edu

**Abstract**

A new algebraic method is developed to reduce the size of the transfer matrix of Ising and three-state Potts ferromagnets, on strips of width *r* sites of square and triangular lattices. This size reduction has been set up in such a way that the maximum eigenvalues of both the reduced and original transfer matrices became exactly the same. In this method we write the original transfer matrix in a special blocked form in such a way that the sums of row elements of a block of the original transfer matrix be the same. The reduced matrix is obtained by replacing each block of the original transfer matrix with the sum of the elements of one of its rows. Our method results in significant matrix size reduction which is a crucial factor in determining the maximum eigenvalue.

## Introduction

There are two general approaches for deriving the partition function of the Ising [1] and Potts [2] models, namely, a combinatorial approach using graphical methods and an algebraic approach using the method of the transfer matrix. Although at zero magnetic field, there is an exact solution for the 2-dimensional Ising model, however, there is no such a solution for the 2-D Potts model. In the absence of an exact solution, series expansions which are based on the graph theory, remain one of the most useful tools in the investigation of the partition function and the critical properties of a model system. For a historical review of the series expansion see Domb [3, 4] and Wu [5] and Biggs [6, 7]. In addition to the complex calculation, the disadvantage of this method is the precision of the calculated critical data which depends on the number of terms in the truncated series.

Among the many various methods for deriving the partition function of the 2-D Ising model, the transfer matrix method is the original and oldest approach. Exact expressions for the eigenvalue of the transfer matrix are available for all 2-D Ising lattices, namely square [8, 9], triangular [10] and honeycomb lattices [11,12]. However, there is no such an expression for the 2-D Potts model. In a recent work, Ranjbar and Parsafar [13] used a finite-width lattice to set up the transfer matrix for the square lattice of the Ising model. The resultant matrix was solved for the model with the width of equal or less than 10 sites. Since the order of the transfer matrix becomes too large for the Potts model, the method cannot be extended to drive the partition function analytically. This method was extended by Gheami et al. [14] for numerical calculation of the critical temperature in the 2-D Ising and the 2-D three-state Potts and 3-D Ising models by the parallel computation method. The numerical methods which are based on the transfer matrix are limited to the size of the matrix.

Finite-size scaling (FSS) has bcome increasingly important in the study of critical phenomena [15, 16]. This is partly due to the progress in theoretical understanding of finite –size effects, and partly due to the application of FSS in the analysis of results from simulation methods. FSS allows us to extract some properties of the infinite system near a phase transition by studying finite, numerically accessible samples. Nightingale [17] has shown how the transfer matrix methods [18-20]could be made more powerful by combining them with FSS to extrapolate their results for finite systems to the thermodynamic limit.

In the present paper we introduce a method for reducing the size of the transfer matrix, in such a way that the maximum eigenvalue ($\lambda_{\max}$) remains the same for both the reduced and original transfer matrices. We have developed an algebraic method on strip of finite width square Ising lattice, to write the original transfer matrix in a special blocked form. In this form the sums of row elements in each block are equal. The reduced transfer matrix is obtained by replacing the blocks of the original transfer matrix with the sum of row elements of that block. The $\lambda_{\max}$ of both the reduced and original transfer matrices have been calculated at the critical point for lattice models with different width. It is observed that the $\lambda_{\max}$ for both the reduced and original transfer matrices are identical.

We have then extended the method to the two-dimensional three-state Potts model on the strip of finite width square lattice. We have shown that our method can be easily used to obtain $\lambda_{\max}$ for the finite width square Potts model from the reduced transfer matrix. The extensions of our method to calculate $\lambda_{\max}$, on the strip of finite width triangular Ising and Potts lattices, are given in Appendices B and C.

## Two-Dimensional Ising Model

Consider a square lattice with the periodic boundary condition composed of slices, each with $p$ rows, where each row has $r$ sites. Each slice has then $p \times r = N$ sites and the coordination number of all sites are the same. In the 2-D Ising model, for any site we define a spin variable $\sigma(i, j) = \pm 1$, in such a way that $i$=1,..., $r$ and $j$=1,..., $p$. We may include the periodic boundary condition as:

$$\sigma(i+r, j) = \sigma(i, j) \tag{1}$$

$$\sigma(i, j+p) = \sigma(i, j) \tag{2}$$

We have only taken into account the interactions among the nearest neighbors. The configurational energy for the model is given as,

$$E(\sigma) = -j\sum_{i=1}^{r}\sum_{j=1}^{p}\{\sigma(i, j)\sigma(i+1, j) + \sigma(i, j)\sigma(i, j+1)\} \tag{3}$$

The canonical partition function, Z ($K$), is

$$Z(K) = \sum_{\{\sigma\}} e^{\frac{-E(\sigma)}{kT}} \tag{4}$$

By substitution of Eq.3 into Eq.4, the canonical partition function can be written as follow [3, 14] :

$$Z(K) = \mathrm{Tr}(\boldsymbol{B}^{p}) \tag{5}$$

where $\boldsymbol{B}$ is the transfer matrix and Tr denotes the trace. To compute the elements of the transfer matrix, consider a matrix $\boldsymbol{A_r}$ with $r$ columns and $n = 2^r$ rows such that each row of the matrix is one of the permutation of 1 and -1 in $r$ columns. The element $a_{i,j} = \pm 1$ of this matrix can be viewed as a spin variable. Therefore, the elements of the transfer matrix $\boldsymbol{B_r}$ can be computed by [3, 14],

$$b_{i,j} = \exp\{K\sum_{k=1}^{r,*}[a_{i,k}a_{i,k+1} + a_{i,k}a_{j,k}]\} \tag{6}$$

where $*$ indicates the periodic boundary condition ($a_{i,k+r} = a_{i,k}$) and $K$ is a constant. The rows of matrix $\boldsymbol{A_r}$ can be grouped into blocks in such a way that the configurational energies, $E_c^{(l)}$ and $E_c^{\circ}$, with $l = 2...r-2$, for each row of the block become the same. We define $E_c^{(l)}$ and $E_c^{\circ}$ as;

$$E_c^{\circ} = \exp\{K\sum_{k=1}^{r,*} a_{i,k}a_{i,k+l} + H\sum_{k=1}^{r} a_{i,k}\} \tag{7}$$

$$E_c^{(l)} = \exp\{K\sum_{k=1}^{r,*} a_{i,k}a_{i,k+l}\} \quad \text{for } l = 2...r-2 \tag{8}$$

where $K$ and $H$ are constant. By this definition, $E_c^{(0)}$ is the nearest neighbor interaction between spins, in the presence of the magnetic field, and $E_c^{(2)}$ is the next nearest neighbor interaction, and $E_c^{(r-2)}$ is the interaction energy between two spins that are separated by $r$-3 spins in each row. Now the matrix $\boldsymbol{A_r}$ can be divided into $g$ blocks. Then the elements of the block $k$ can be represented by $a_{i,j}^{(k)}$ with $i = 1...m_{(k)}$, $j = 1...r$, and $k = 1...g$, where $m_k$ is the number of rows in the block $k$. With this definition, each block has the following characteristics:

(a)The number of +1 (-1) spins in each row of a block are the same, (b) The rows of a block are the cyclic permutation of each other, (c) The first block has one row with all elements equal to $-1$, and the block $g$ has one row with all elements equal to +1. For example let us consider the case where $r$ = 4. There are $2^4$ different states in which spins can be arranged in 4 sites. We can show these states by the following $2^4 \times 4$ matrix,

$$
\boldsymbol{A}_4 = \left( \begin{array}{cccc}
-1 & -1 & -1 & -1 \\ \hdashline
1 & -1 & -1 & -1 \\
-1 & 1 & -1 & -1 \\
-1 & -1 & 1 & -1 \\
-1 & -1 & -1 & 1 \\ \hdashline
1 & 1 & -1 & -1 \\
-1 & 1 & 1 & -1 \\
1 & -1 & -1 & 1 \\
-1 & -1 & 1 & 1 \\ \hdashline
1 & -1 & 1 & -1 \\
-1 & 1 & -1 & 1 \\ \hdashline
1 & 1 & 1 & -1 \\
1 & 1 & -1 & 1 \\
1 & -1 & 1 & 1 \\
-1 & 1 & 1 & 1 \\ \hdashline
1 & 1 & 1 & 1
\end{array} \right) \tag{9}
$$

where each row represents one of the possible spin configurations of 4 sites. As shown in Eq.9, the matrix $\boldsymbol{A_4}$ is devided into 6 blocks, and the rows of each block are the cyclic permutation of each other.

Then by using such a notation, the elements of the transfer matrix $\boldsymbol{B_r}$ can be represented as follow:

$$
b_{i,j}^{(t),(s)} = \exp\{K \sum_{k=1}^{r,*} [a_{i,k}^{(t)} a_{i,k+1}^{(t)} + a_{i,k}^{(t)} a_{j,k}^{(s)}]\} \tag{10}
$$

or

$$
b_{i,j}^{(t),(s)} = h_i^{(t)} f_{i,j}^{(t),(s)} \tag{11}
$$

where, $h_i^{(t)}$ represents the interactions among spins of row *i* of the block *t* of the matrix $\boldsymbol{A_r}$, and is defined as,

$$
h_i^{(t)} = \exp\{K \sum_{k=1}^{r,*} a_{i,k}^{(t)} a_{i,k+1}^{(t)}\} \tag{12}
$$

and, $f_{i,j}^{(t),(s)}$ represents the column-column interactions among spins of row *i* of the block *t* and row *j* of the block *s* of the matrix $\boldsymbol{A_r}$, and is defined as,

$$
f_{i,j}^{(t),(s)} = \exp\{K \sum_{k=1}^{r} a_{i,k}^{(t)} a_{j,k}^{(s)}\} \tag{13}
$$

Now the transfer matrix $\boldsymbol{B_r}$ can be divided into $g \times g$ blocks, in which the block $\boldsymbol{B}^{(t),(s)}$ has $m_{(t)}$ rows and $m_{(s)}$ columns. Thus we may represent the transfer matrix $\boldsymbol{B_r}$ as;

$$\boldsymbol{B_r} = \left| \begin{array}{c|c|c|c} B^{(1),(1)} & B^{(1),(2)} & \cdots & B^{(1),(g)} \\ \hline B^{(2),(1)} & B^{(2),(2)} & \cdots & B^{(2),(g)} \\ \hline \vdots & \vdots & \ddots & \vdots \\ \hline B^{(g),(1)} & B^{(g),(2)} & \cdots & B^{(g),(g)} \end{array} \right| \tag{14}$$

In which, each block has the following characteristics:

(1) We define operator $\hat{\mathbf{P}}^{(s)}$ as a permutation operator that converts one row of block $s$ to another row of this block for the matrix $\boldsymbol{A_r}$. So we can write,

$$\hat{\mathbf{P}}^{(s)} b_{i,j}^{(t),(s)} = h_i^{(t)} \hat{\mathbf{P}}^{(s)} f_{i,j}^{(t),(s)} = h_i^{(t)} f_{i,k}^{(t),(s)} = b_{i,k}^{(t),(s)} \tag{15}$$

and similarly,

$$\hat{\mathbf{P}}^{(t)} b_{i,j}^{(t),(s)} = \hat{\mathbf{P}}^{(t)} h_i^{(t)} f_{i,j}^{(t),(s)} = h_q^{(t)} f_{q,j}^{(t),(s)} = b_{q,j}^{(t),(s)} \tag{16}$$

The function $h_i^{(t)}$ is equal to $h_q^{(t)}$ (see properties (a) and (b) of matrix $\boldsymbol{A_r}$) and the function $f_{i,k}^{(t),(s)}$ is obtained from function $f_{i,j}^{(t),(s)}$ by replacing the row $j$ with the row $k$ of the block $s$ of the matrix $\boldsymbol{A_r}$. Therefore instead of the cyclic permutation on the row $j$ of the block $s$ of the matrix $\boldsymbol{A_r}$, we can obtain the same result, by replacing the row $i$ with row $q$ of the block $t$, or cyclic permutation on the row $i$ of the block $t$ of the matrix $\boldsymbol{A_r}$ (see properties (a) and (b) of matrix $\boldsymbol{A_r}$), or

$$\hat{\mathbf{P}}^{(s)} b_{i,j}^{(t),(s)} = \hat{\mathbf{P}}^{(t)} b_{i,j}^{(t),(s)} \tag{17}$$

By using Eq.15, we may write,

$$\sum_{j=1}^{m_{(s)}} b_{i,j}^{(t),(s)} = \hat{\mathbf{P}}^{(s)} \sum_{j=1}^{m_{(s)}} b_{i,j}^{(t),(s)} \tag{18}$$

This is due to the fact that the effect of operation $\hat{\mathbf{P}}^{(s)}$ is to convert the terms in the summand to each other. Substituting Eq.17 into Eq.18, we obtain,

$$\hat{\mathbf{P}}^{(s)} \sum_{j=1}^{m_{(s)}} b_{i,j}^{(t),(s)} = \hat{\mathbf{P}}^{(t)} \sum_{j=1}^{m_{(s)}} b_{i,j}^{(t),(s)} = \sum_{j=1}^{m_{(s)}} b_{q,j}^{(t),(s)} \tag{19}$$

or equivalently

$$\sum_{j=1}^{m_{(s)}} b_{1,j}^{(t),(s)} = \sum_{j=1}^{m_{(s)}} b_{2,j}^{(t),(s)} = \ldots = \sum_{j=1}^{m_{(s)}} b_{m_{(t)},j}^{(t),(s)} = \alpha_{t,s} \tag{20}$$

Therefore, the sums of row elements of the block $\boldsymbol{B}^{(t),(s)}$ of the matrix $\boldsymbol{B_r}$ are equal.

2) With the same reasoning which led to Eq.20, it can be easily shown that,

$$\sum_{i=1}^{m_{(t)}} b_{i,1}^{(t),(s)} = \sum_{i=1}^{m_{(t)}} b_{i,2}^{(t),(s)} = \ldots = \sum_{i=1}^{m_{(t)}} b_{i,m_{(s)}}^{(t),(s)} = \beta_{t,s} \tag{21}$$

For example a sample transfer matrix $\boldsymbol{B_4}$ whose elements are computed by applying Eq.10 to the matrix $\boldsymbol{A_4}$ (Eq.9) becomes:

$$\boldsymbol{B}_4 = \left[\begin{array}{c|cccc|cccc|cc|cccc|c}
m^8 & m^6 & m^6 & m^6 & m^6 & m^4 & m^4 & m^4 & m^4 & m^4 & m^4 & m^2 & m^2 & m^2 & m^2 & 1 \\ \hline
m^2 & m^4 & 1 & 1 & 1 & m^2 & 1/m^2 & m^2 & 1/m^2 & m^2 & 1/m^2 & 1 & 1 & 1 & 1/m^4 & 1/m^2 \\
m^2 & 1 & m^4 & 1 & 1 & m^2 & m^2 & 1/m^2 & 1/m^2 & 1/m^2 & m^2 & 1 & 1 & 1/m^4 & 1 & 1/m^2 \\
m^2 & 1 & 1 & m^4 & 1 & 1/m^2 & m^2 & 1/m^2 & m^2 & m^2 & 1/m^2 & 1 & 1/m^4 & 1 & 1 & 1/m^2 \\
m^2 & 1 & 1 & 1 & m^4 & 1/m^2 & 1/m^2 & m^2 & m^2 & 1/m^2 & m^2 & 1/m^4 & 1 & 1 & 1 & 1/m^2 \\ \hline
1 & m^2 & m^2 & 1/m^2 & 1/m^2 & m^4 & 1 & 1 & 1/m^4 & 1 & 1 & m^2 & m^2 & 1/m^2 & 1/m^2 & 1 \\
1 & 1/m^2 & m^2 & m^2 & 1/m^2 & 1 & m^4 & 1/m^4 & 1 & 1 & 1 & m^2 & 1/m^2 & 1/m^2 & m^2 & 1 \\
1 & m^2 & 1/m^2 & 1/m^2 & m^2 & 1 & 1/m^4 & m^4 & 1 & 1 & 1 & 1/m^2 & m^2 & m^2 & 1/m^2 & 1 \\
1 & 1/m^2 & 1/m^2 & m^2 & m^2 & 1/m^4 & 1 & 1 & m^4 & 1 & 1 & 1/m^2 & 1/m^2 & m^2 & m^2 & 1 \\ \hline
1/m^4 & 1/m^2 & 1/m^6 & 1/m^2 & 1/m^6 & 1/m^4 & 1/m^4 & 1/m^4 & 1/m^4 & 1 & 1/m^8 & 1/m^2 & 1/m^6 & 1/m^2 & 1/m^6 & 1/m^4 \\
1/m^4 & 1/m^6 & 1/m^2 & 1/m^6 & 1/m^2 & 1/m^4 & 1/m^4 & 1/m^4 & 1/m^4 & 1/m^8 & 1 & 1/m^6 & 1/m^2 & 1/m^6 & 1/m^2 & 1/m^4 \\ \hline
1/m^2 & 1 & 1 & 1 & 1/m^4 & m^2 & m^2 & 1/m^2 & 1/m^2 & m^2 & 1/m^2 & m^4 & 1 & 1 & 1 & m^2 \\
1/m^2 & 1 & 1 & 1/m^4 & 1 & m^2 & 1/m^2 & m^2 & 1/m^2 & 1/m^2 & m^2 & 1 & m^4 & 1 & 1 & m^2 \\
1/m^2 & 1 & 1/m^4 & 1 & 1 & 1/m^2 & 1/m^2 & m^2 & m^2 & m^2 & 1/m^2 & 1 & 1 & m^4 & 1 & m^2 \\
1/m^2 & 1/m^4 & 1 & 1 & 1 & 1/m^2 & m^2 & 1/m^2 & m^2 & 1/m^2 & m^2 & 1 & 1 & 1 & m^4 & m^2 \\ \hline
1 & m^2 & m^2 & m^2 & m^2 & m^4 & m^4 & m^4 & m^4 & m^4 & m^4 & m^6 & m^6 & m^6 & m^6 & m^8
\end{array}\right] \tag{22}$$

where, $m$ is defined by,

$$m = \exp\{K\} \tag{23}$$

Computation of the eigenvalues and the corresponding eigenvectors of the matrix $\boldsymbol{B_r}$ (Eq.14) requires the solution of the following set of homogenous equations;

$$\begin{aligned}
&(b_{1,1}^{(1),(1)} - \lambda)x_1^{(1)} + b_{1,1}^{(1),(2)}x_1^{(2)} + b_{1,2}^{(1),(2)}x_2^{(2)} + \ldots + b_{1,1}^{(1),(g)}x_1^{(g)} = 0 \\
&b_{1,1}^{(2),(1)}x_1^{(1)} + (b_{1,1}^{(2),(2)} - \lambda)x_1^{(2)} + b_{1,2}^{(2),(2)}x_2^{(2)} + \ldots + b_{1,1}^{(2),(g)}x_1^{(g)} = 0 \\
&b_{2,1}^{(2),(1)}x_1^{(1)} + b_{2,1}^{(2),(2)}x_1^{(2)} + (b_{2,2}^{(2),(2)} - \lambda)x_2^{(2)} + \ldots + b_{1,1}^{(2),(g)}x_1^{(g)} = 0 \\
&\vdots \\
&b_{1,1}^{(g),(1)}x_1^{(1)} + b_{1,1}^{(g),(2)}x_1^{(2)} + b_{1,2}^{(g),(2)}x_2^{(2)} + \ldots + (b_{1,1}^{(g),(g)} - \lambda)x_1^{(g)} = 0
\end{aligned} \tag{24}$$

For each block (e.g. block $s$) consider a special case where,

$$x_1^{(s)} = x_2^{(s)} = \ldots = x_{m_s}^{(s)} = x^{(s)} \tag{25}$$

By inserting Eq.25 into Eqs.24 and using Eq.20, the set of $n$ homogenous equations (Eqs.24) will be converted into the following set of equations;

$$\begin{aligned}
&(\alpha_{1,1} - \lambda)x^{(1)} + \alpha_{1,2}x^{(2)} + \ldots + \alpha_{1,g}x^{(g)} = 0 \\
&\alpha_{2,1}x^{(1)} + (\alpha_{2,2} - \lambda)x^{(2)} + \ldots + \alpha_{2,g}x^{(g)} = 0 \\
&\alpha_{g,1}x^{(1)} + \alpha_{g,2}x^{(2)} + \ldots + (\alpha_{g,g} - \lambda)x^{(g)} = 0
\end{aligned} \tag{26}$$

By solving Eqs.26, we obtain $g$ different eigenvalues $\lambda$ and corresponding eigenvectors that are also the solutions of Eqs.24, but Eqs.24 have some solutions that do not satisfy the condition of Eq.25.

However, we have observed that the maximum eigenvalue ($\lambda_{\max}$) of the $g$th order matrix $\alpha_4$ with elements $\alpha_{i,j}$ is equal to $\lambda_{\max}$ of the $n$th order matrix $\boldsymbol{B_r}$ with the elements $b_{i,j}^{(t),(s)}$. This means that $\lambda_{\max}$ is the same for both $\boldsymbol{B_r}$ and $\alpha_r$ matrices. For example the reduced matrix $\alpha_4$ which is computed from the original transfer matrix $\boldsymbol{B_4}$ (Eq.22) can be written as:

$$\boldsymbol{\alpha}_4 = \begin{pmatrix} m^8 & 4m^6 & 4m^4 & 2m^4 & 4m^2 & 1 \\ m^2 & 3+m^4 & \frac{2}{m^2}+2m^2 & \frac{1}{m^2}+m^2 & \frac{1}{m^4}+3 & \frac{1}{m^2} \\ 1 & \frac{2}{m^2}+2m^2 & \frac{1}{m^4}+2+m^4 & 2 & \frac{2}{m^2}+2m^2 & 1 \\ \frac{1}{m^4} & \frac{2}{m^6}+\frac{2}{m^2} & \frac{4}{m^4} & \frac{1}{m^8}+1 & \frac{2}{m^6}+\frac{2}{m^2} & \frac{1}{m^4} \\ \frac{1}{m^2} & \frac{1}{m^4}+3 & \frac{2}{m^2}+2m^2 & \frac{1}{m^2}+m^2 & 3+m^4 & m^2 \\ 1 & 4m^2 & 4m^4 & 2m^4 & 4m^6 & m^8 \end{pmatrix} \tag{27}$$

The eigenvalues of both $\alpha_4$ and $\boldsymbol{B_4}$ matrices (Eq.27, Eq.22) are given in the appendix A. Size reduction of the transfer matrix can be easily applied to the triangular Ising lattice (see appendix B). Such a reduction becomes important when *r* has large values (namely, larger than 12). In these cases the size of the transfer matrix is larger than $2^{12} \times 2^{12}$ and hence the computation of the largest eigenvalue ($\lambda_{\max}$) is difficult. However the size of the reduced transfer matrix is much smaller than that of the original transfer matrix $\boldsymbol{B_r}$, in such a way that the $\lambda_{\max}$ can be easily calculated from the reduced matrix (see Table 1). The reduced free energy per site ($(1/r)\ln\lambda_{\max}$)for the square Ising model at the critical point has been calculated for both the reduced and original transfer matrices with different lattice sizes. Such a calculation has also been done on the strip of the triangular Ising lattice, and the results for the reduced and original transfer matrices are compared in Table 2. As it is shown in Table 2, the results for both matrices are identical.

**Two-Dimensional Three-State Potts Model**

Consider a square lattice with the periodic boundary condition, composed of slices, each with *p* rows, where each row has *r* sites. Then each slice has $p \times r = N$ sites and the coordination number of all sites are the same. For any site we define a spin variable $\sigma(i, j) = 0,1,2$, so that $i$=1,..., $r$ and $j$=1,..., $p$. The configurational energy of the standard 3-state Potts model is given by [5],

$$E(\sigma) = \sum_{i=1}^{r}\sum_{j=1}^{p} - J[\delta_{\sigma(i,j),\sigma(i,j+1)} + \delta_{\sigma(i,j),\sigma(i+1,j)}] \tag{28}$$

where

$$\delta_{i,j} = 1 \quad \text{for} \quad i = j$$

$$\delta_{i,j} = 0 \quad \text{for} \quad i \neq j \tag{29}$$

The canonical partition function, Z ($K$), is given by [5, 14],

$$Z(K) = \mathrm{Tr}\,(\boldsymbol{C_r}^{\,p}) \tag{30}$$

where $\boldsymbol{C_r}$ is the transfer matrix. To compute the elements of the transfer matrix, consider matrix $\boldsymbol{D_r}$ with $r$ columns and $n = 3^r$ rows that each row of the matrix is one of the permutation of the 0, 1, and 2 in the $r$ columns. The element $d_{i,j}$ =0, 1, and 2, of this matrix can be viewed as a spin variable. The rows of matrix $\boldsymbol{D_r}$ can be blocked in such a way that the configurational energies, $E^{(l)}_{\mathrm{c,Potts}}$ and $E^{\circ}_{\mathrm{c,Potts}}$, with $l = 2...r-2$, for each row of the block becomes the same. We define $E^{(l)}_{\mathrm{c,Potts}}$ and $E^{\circ}_{\mathrm{c,Potts}}$ as;

$$E^{\circ}_{\mathrm{c,Potts}} = K\sum_{k=1}^{r,*}\delta_{d_{i,k},d_{i,k+l}} + H\sum_{k=1}^{r}\delta_{d_{i,k},0} \tag{31}$$

$$E^{(l)}_{\mathrm{c,Potts}} = K\sum_{k=1}^{r,*}\delta_{d_{i,k},d_{i,k+l}} \quad \text{for } l = 2...r-2, \tag{32}$$

where $K$ and $H$ are constants. In this definition $E^{\circ}_{\mathrm{c,Potts}}$ is the nearest neighbor interactions among spins in each row, and $E^{(2)}_{\mathrm{c,Potts}}$ is the next nearest neighbor interactions, and $E^{(r-2)}_{\mathrm{c,Potts}}$ is the interaction energy between two spins that are separated by $r$-3 spins in each row. Now the matrix $\boldsymbol{D_r}$ can be divided into $g$ blocks. Then the elements of the block $k$ can be represented by $d_{i,j}^{(k)}$ with $i = 1...m_{(k)}$, $j = 1...r$, and $k = 1...g$, where $m_k$ is the number of rows in the block $k$. With such a definition, each block will have the following characteristics:

(a) Each row of a block has the same number of spins with 0 values. (b) The rows of a block which have the same number of spins with 0, 1, and 2, are cyclic permutation of each other. (c) One half of the rows of a block can be obtained from the other half by replacing spin 1 with spin 2 and vice versa. (d) Each matrix has a block that has one row with all elements equal to 0, and another block with two rows, one row with all elements equal to +1 and one row with all elements equal to -1. For instance, consider the case for which $r$ = 3. There are $3^3$ different states that spins 0, 1, and 2 which can be arranged in three sites. We can show these states by the following $3^3 \times 3$ matrix $\boldsymbol{D_3}$,

$$\boldsymbol{D_3}=\left[\begin{array}{ccc} 0 & 0 & 0 \\ \hline 1 & 0 & 0 \\ 2 & 0 & 0 \\ 0 & 1 & 0 \\ 0 & 2 & 0 \\ 0 & 0 & 1 \\ 0 & 0 & 2 \\ \hline 1 & 1 & 0 \\ 2 & 2 & 0 \\ 1 & 0 & 1 \\ 0 & 1 & 1 \\ 2 & 0 & 2 \\ 0 & 2 & 2 \\ \hline 2 & 1 & 0 \\ 1 & 2 & 0 \\ 2 & 0 & 1 \\ 0 & 2 & 1 \\ 1 & 0 & 2 \\ 0 & 1 & 2 \\ \hline 1 & 1 & 1 \\ 2 & 2 & 2 \\ \hline 2 & 1 & 1 \\ 1 & 2 & 1 \\ 2 & 2 & 1 \\ 1 & 1 & 2 \\ 2 & 1 & 2 \\ 1 & 2 & 2 \end{array}\right] \quad (33)$$

As shown by Eq.33, in this case matrix $\boldsymbol{D_3}$ has 6 blocks. Then by this notation, the elements of the transfer matrix $\boldsymbol{C_r}$ can be represented as follows [5, 14];

$$c_{i,j}^{(t),(s)} = \exp\{K\sum_{k=1}^{r,*}[\delta_{d_{i,k}^{(t)},d_{i,k+1}^{(t)}} + \delta_{d_{i,k}^{(t)},d_{j,k}^{(s)}}]\} \tag{34}$$

Now the transfer matrix $\boldsymbol{C_r}$ can be presented by a matrix with $g \times g$ blocks (Eq.35), in which the block $\boldsymbol{C}^{(t),(s)}$ has $m_{(t)}$ rows and $m_{(s)}$ columns.

$$\boldsymbol{C_r} = \left|\begin{array}{c|c|c|c} C^{(1),(1)} & C^{(1),(2)} & \cdots & C^{(1),(g)} \\ \hline C^{(2),(1)} & C^{(2),(2)} & \cdots & C^{(2),(g)} \\ \hline \vdots & \vdots & \ddots & \vdots \\ \hline C^{(g),(1)} & C^{(g),(2)} & \cdots & C^{(g),(g)} \end{array}\right| \tag{35}$$

Each block of $\boldsymbol{C_r}$ has the following characteristics:

(1) As before, the elements of the transfer matrix $\boldsymbol{C_r}$ can be written into two terms;

$$c_{i,j}^{(t),(s)} = y_i^{(t)} g_{i,j}^{(t),(s)} \tag{36}$$

where, $y_i^{(t)}$ represents the interactions among spins of row *i* of the block *t* of the matrix $\boldsymbol{D_r}$, and is defined by,

$$y_i^{(t)} = \exp\{K\sum_{k=1}^{r,*}\delta_{d_{i,k}^{(t)},d_{i,k+1}^{(t)}}\} \tag{37}$$

and $g_{i,j}^{(t),(s)}$ represents the column-column interactions among spins of row *i* of the block *t* and row *j* of the block *s*, of the matrix $\boldsymbol{D_r}$, and is defined by,

$$g_{i,j}^{(t),(s)} = \exp\{K\sum_{k=1}^{r}\delta_{d_{i,k}^{(t)},d_{j,k}^{(s)}}\} \tag{38}$$

We define operator $\hat{\mathbf{P}}^{(s)}$ as a permutation operator that converts one row of block *s* to another row of this block of the matrix $\boldsymbol{D_r}$. This conversion can be done in two ways: By cyclic permutation or by converting 1 to 2 and vice versa. So we can write,

$$\hat{\mathbf{P}}^{(s)} c_{i,j}^{(t),(s)} = y_i^{(t)} \hat{\mathbf{P}}^{(s)} g_{i,j}^{(t),(s)} = y_i^{(t)} g_{i,k}^{(t),(s)} = c_{i,k}^{(t),(s)} \qquad (39)$$

and similarly,

$$\hat{\mathbf{P}}^{(t)} c_{i,j}^{(t),(s)} = \hat{\mathbf{P}}^{(t)} y_i^{(t)} g_{i,j}^{(t),(s)} = y_q^{(t)} g_{q,j}^{(t),(s)} = c_{q,j}^{(t),(s)} \qquad (40)$$

The function $y_i^{(t)}$ is equal $y_q^{(t)}$ (see properties a, b, and c of matrix $\boldsymbol{D_r}$) and the function $g_{i,k}^{(t),(s)}$ is obtained from function $g_{i,j}^{(t),(s)}$ by replacing the row *j* with the row *k* of the block *s* of the matrix $\boldsymbol{D_r}$. So instead of permutation on the row *j* of the block *s* of the matrix $\boldsymbol{D_r}$, we can obtain the same result, by replacing row *i* with row *q*, of the block *t*, or by permutation on the row *i* of the block *t* for the matrix $\boldsymbol{D_r}$ (see properties a, b, and c of the matrix $\boldsymbol{D_r}$), hence

$$\hat{\mathbf{P}}^{(s)} c_{i,j}^{(t),(s)} = \hat{\mathbf{P}}^{(t)} c_{i,j}^{(t),(s)} \qquad (41)$$

By using Eq.39 we may write,

$$\sum_{j=1}^{m_{(s)}} c_{i,j}^{(t),(s)} = \hat{\mathrm{P}}^{(s)} \sum_{j=1}^{m_{(s)}} c_{i,j}^{(t),(s)} \qquad (42)$$

This is due to the fact that the effect of operation $\hat{\mathbf{P}}^{(s)}$ is to convert the terms in the summand to each other. By inserting Eq.41 into Eq.42, we can write,

$$\hat{\mathrm{P}}^{(s)} \sum_{j=1}^{m_{(s)}} c_{i,j}^{(t),(s)} = \hat{\mathrm{P}}^{(t)} \sum_{j=1}^{m_{(s)}} c_{i,j}^{(t),(s)} = \sum_{j=1}^{m_{(s)}} c_{q,j}^{(t),(s)} \qquad (43)$$

or equivalently

$$\sum_{j=1}^{m_{(s)}} c_{1,j}^{(t),(s)} = \sum_{j=1}^{m_{(s)}} c_{2,j}^{(t),(s)} = \ldots = \sum_{j=1}^{m_{(s)}} c_{m_{(t)},j}^{(t),(s)} = \gamma_{t,s} \qquad (44)$$

So the sum of elements of each row of the block $\boldsymbol{C}^{(t),(s)}$ for the matrix $\boldsymbol{C_r}$ is the same.

2) With the same reasoning which led to Eq.44, it can be easily shown that;

$$\sum_{i=1}^{m_{(t)}} c_{i,1}^{(t),(s)} = \sum_{i=1}^{m_{(t)}} c_{i,2}^{(t),(s)} = \ldots = \sum_{i=1}^{m_{(t)}} c_{i,m_{(s)}}^{(t),(s)} = \chi_{t,s} \tag{45}$$

Computation of eigenvalues and the corresponding eigenvectors of the matrix $\boldsymbol{C_r}$ are similar to that of matrix $\boldsymbol{B_r}$ (Eq.24). By inserting the condition of Eq.25, we can calculate the maximum eigenvalue ($\lambda_{\max}$) of the matrix $\boldsymbol{C_r}$ from its corresponding reduced matrix $\boldsymbol{\Gamma_r}$ with the element $\gamma_{i,j}$ given by Eq.44. For example the reduced matrix $\boldsymbol{\Gamma}_3$ that computed from the original matrix $\boldsymbol{C_3}$ for the case with $r$ =3 can be written as:

$$\boldsymbol{\Gamma}_3 = \begin{bmatrix} m^6 & 6m^5 & 6m^4 & 6m^4 & 2m^3 & 6m^3 \\ m^3 & 4m^2+m^3+m^4 & 2m+2m^2+2m^3 & 2m+2m^2+2m^3 & m+m^2 & 3m+3m^2 \\ m^2 & 2m+2m^2+2m^3 & 2m+3m^2+m^4 & 2m+2m^2+2m^3 & m+m^3 & m+4m^2+m^3 \\ m & 2+2m^2+2m & 2+2m^2+2m & 3m+2+m^3 & 2m & 2+2m^2+2m \\ m^3 & 3m^3+3m^4 & 3m^3+3m^5 & 6m^4 & m^3+m^6 & 3m^4+3m^5 \\ m & 3m+3m^2 & m+4m^2+m^3 & 2m+2m^2+2m^3 & m^2+m^3 & m+2m^3+2m^2+m^4 \end{bmatrix} \tag{46}$$

Size reduction of the transfer matrix can be easily applied to the triangular Potts model (see appendix C). Again, such size reduction of the transfer matrix becomes important when $r$ is large (namely, larger than 8). In these cases, the size of the original transfer matrix is larger than $3^8 \times 3^8$ and computation of its largest eigenvalue ($\lambda_{\max}$) is difficult, however the size of its reduced transfer matrix $\boldsymbol{\Gamma_r}$ may be so small (compare to the original transfer matrix $\boldsymbol{C_r}$ ) that the calculation of $\lambda_{\max}$ may be easily done (See Table 3). The reduced free energy per site ($(1/r)\ln\lambda_{\max}$) for the square 3-state Potts model at the critical point has been calculated for both the reduced and original transfer matrices with different lattice sizes. Such a calculation was also been done for the triangular 3-state Potts model, and the results are compared with the values obtained by de Queiroz [21] in Tables 4 and 5. As it is seen from Tables 4 and 5, the results for the reduced transfer matrix are exactly the same as those of the original transfer matrix.

## Conclusion

We have shown that our algebraic method for the size reduction of the transfer matrix, can principally be applied to the 2-dimensional Ising and 2-dimensional three-state Potts models. In fact our result ($\lambda_{\max}$) for the reduced transfer matrix is identical to that calculated from the original transfer matrix for each model (see Tables 2 and 4).

Although at present a wide variety of mathematical software are available, which can be used for analytic computation of the maximum eigenvalue of the transfer matrix, all of them are limited to cases with small matrix size. Therefore the size reduction technique for the transfer matrix is still important for such calculations. For numerical investigation on finite-size lattices, our method of size reduction for the transfer matrix can be used for cases with larger lattices (see Tables 1, 4, and 5).

Although, an explicit mathematical proof for the equality of maximum eigenvalues of both the reduced and original transfer matrices is not given, for the 2-D Ising and Potts models at least, we have shown that their $\lambda_{\max}$ are identical.

Finally, we expect our method to be extended to the two dimensional models in the presence of magnetic field. This could be done by adding another condition to Eqs.7 and 8 for partitioning the matrix $\boldsymbol{A_r}$, which is still under investigation.

## Acknowledgment

We acknowledge the Iranian National Research Council for the financial support, and also Dr. M. Ashrafizaadeh for his useful comment.

## Appendix A

In this appendix the results of analytical computation of the eigenvalues of the $\boldsymbol{B_4}$ matrix (Eq.22) are provided,

$$\lambda_1 = \lambda_2 = \lambda_3 = \lambda_4 = \frac{m^8 - 2m^4 + 1}{m^4} \tag{A.1}$$

$$\lambda_5 = \lambda_6 = \frac{m^8 + 2m^6 - 2m^2 - 1}{m^4} \tag{A.2}$$

$$\lambda_7 = \lambda_8 = \frac{m^8 - 2m^6 + 2m^2 - 1}{m^4} \tag{A.3}$$

$$\lambda_9 = \frac{(m^8 + 2m^4 + 1 + \sqrt{m^{16} + 14m^8 + 1})(m^2 + 1)(m + 1)(m - 1)}{2m^8} \tag{A.4}$$

$$\lambda_{10} = \frac{(m^8 + 2m^4 + 1 + \sqrt{m^{16} + 14m^8 + 1})(m^2 + 1)(m + 1)(m - 1)}{2m^4} \tag{A.5}$$

$$\lambda_{11} = \frac{(m^8 + 2m^4 + 1 - \sqrt{m^{16} + 14m^8 + 1})(m^2 + 1)(m + 1)(m - 1)}{2m^8} \tag{A.6}$$

$$\lambda_{12} = \frac{(m^8 + 2m^4 + 1 - \sqrt{m^{16} + 14m^8 + 1})(m^2 + 1)(m + 1)(m - 1)}{2m^4} \tag{A.7}$$

$$\lambda_{13} = \frac{-1}{4}(-1 - m^{16} - 2m^{12} - 10m^8 - 2m^4 + \sqrt{p} - g(p,m))/m^8 \tag{A.8}$$

where,

$$g(p,m) = 2^{\frac{1}{2}}(p - \sqrt{p}(1 + 10m^8 + m^{16} + 2m^{12} + 2m^4) + 28m^{20} - 64m^{16} + 28m^{12} + 4m^{28} + 4m^4)^{\frac{1}{2}} \tag{A.10}$$

and

$$p = m^{32} - 4m^{28} + 8m^{24} + 52m^{20} + 142m^{16} + 52m^{12} + 8m^8 - 4m^4 + 1 \tag{A.11}$$

$$\lambda_{14} = \frac{-1}{4}(-1 - m^{16} - 2m^{12} - 10m^8 - 2m^4 + \sqrt{p} + g(p,m))/m^8 \tag{A.12}$$

$$\lambda_{15} = \frac{1}{4}(1 + m^{16} + 2m^{12} + 10m^8 + 2m^4 + \sqrt{p} - f(p,m))/m^8 \tag{A.13}$$

$$\lambda_{\max} = \lambda_{16} = \frac{1}{4}(1+m^{16} + 2m^{12} + 10m^8 + 2m^4 + \sqrt{p} + f(p,m))/m^8 \qquad \text{(A.14)}$$

where

$$f(p,m) = 2^{\frac{1}{2}}(p + \sqrt{p}(1+10m^8 + m^{16} + 2m^{12} + 2m^4) + 28m^{20} - 64m^{16} + 28m^{12} + 4m^{28} + 4m^4)^{\frac{1}{2}}$$

(A.15)

The eigenvalues $\lambda_{10}$ to $\lambda_{16}$ (Eqs.A.6 to A.15)are also eigenvalues of the $\boldsymbol{\alpha}_4$ matrix (Eq.27).Notice that the maximum eigenvalue is the same for both the $\boldsymbol{\alpha}_4$ (Eq.27)and $\boldsymbol{B_4}$ (Eq.22) matrices, even though all eigenvalues are not the same.

**Appendix B**

The elements of the transfer matrix of the triangular Ising model (***T_r***)could be computed by using matrix ***A_r*** as follow [10,14]:

$$t_{i,j}^{(w).(s)} = \exp\{K\sum_{k=1}^{r,*}[a_{i,k}^{(w)}a_{i,k+1}^{(w)} + a_{i,k}^{(w)}a_{j,k}^{(s)} + a_{i,k}^{(w)}a_{j,k+1}^{(s)}]\} \quad \text{(B1)}$$

or equivalently,

$$t_{i,j}^{(w),(s)} = h_i^{(w)} v_{i,j}^{(w),(s)} f_{i,j}^{(w),(s)} \quad \text{(B2)}$$

where $v_{i,j}^{(w),(s)}$ represents the column-column interactions among spins of row *i* of the block *w* and row *j* of the block *s* of matrix ***A_r*** in such a way that the element of column *k* interacts with the element of column *k*+1, which is defined as,

$$v_{i,j}^{(w),(s)} = \exp\{K\sum_{k=1}^{r,*} a_{i,k}^{(w)} a_{j,k+1}^{(s)}\} \quad \text{(B3)}$$

and the functions $h_i^{(w)}$ and $f_{i,j}^{(w),(s)}$ have their preceding meaning (Eqs.12 and 13). Now the transfer matrix ***T_r*** can be divided into $g \times g$ blocks, in which the block $\boldsymbol{T}^{(w),(s)}$ has $m_{(w)}$ rows and $m_{(s)}$ columns. The effect of operator $\hat{\mathbf{P}}^{(s)}$ on the $v_{i,j}^{(w),(s)}$ is the same as its effect on $f_{i,j}^{(w),(s)}$ (Eqs.15 and 16) and with the same reasoning which led to Eq.20, it can be easily shown that,

$$\sum_{j=1}^{m_{(s)}} t_{1,j}^{(t),(s)} = \sum_{j=1}^{m_{(s)}} t_{2,j}^{(w),(s)} = \ldots = \sum_{j=1}^{m_{(s)}} t_{m_{(w)},j}^{(w),(s)} = \phi_{t,s} \quad \text{(B4)}$$

As in the case of the square Ising lattice, the maximum eigenvalue can be calculated from the reduced transfer matrix $\Phi$ with the elements $\phi_{i,j}$.

**Appendix C**

The elements of the transfer matrix on the strip of the triangular three-state Potts model (***W*<sub></sub>**$_r$)could be computed by using matrix $\boldsymbol{D_r}$ as follow [5,14]:

$$w_{i,j}^{(t).(s)} = \exp\{K\sum_{k=1}^{r,*}[\delta_{d_{i,k}^{(t)},d_{i,k+1}^{(t)}} + \delta_{d_{i,k}^{(t)},d_{j,k}^{(s)}} + \delta_{d_{i,k}^{(t)},d_{j,k+1}^{(s)}}]\} \qquad \text{(C1)}$$

or equivalently,

$$w_{i,j}^{(t),(s)} = y_i^{(t)} q_{i,j}^{(t),(s)} g_{i,j}^{(t),(s)} \qquad \text{(C2)}$$

where $q_{i,j}^{(t),(s)}$ represents the column-column interactions among spins of row $i$ of the block $t$ and row $j$ of the block $s$ of matrix $\boldsymbol{A_r}$ in such a way that the element of column $k$ interacts with the element of columns $k$+1, and is defined as,

$$q_{i,j}^{(t),(s)} = \exp\{K\sum_{k=1}^{r} \delta_{d_{i,k}^{(t)},d_{j,k+1}^{(s)}}\} \qquad \text{(C3)}$$

and the functions $y_i^{(t)}$ and $g_{i,j}^{(t),(s)}$ have their preceding meaning (Eqs.37 and 38). Now the transfer matrix $\boldsymbol{W_r}$ can be divided into $g \times g$ blocks, in which the block $\boldsymbol{W}^{(t),(s)}$ has $m_{(t)}$ rows and $m_{(s)}$ columns. The effect of operator $\hat{\mathbf{P}}^{(s)}$ on the $q_{i,j}^{(t),(s)}$ is the same as its effect on $g_{i,j}^{(t),(s)}$ (Eqs.39 and 40) and with the same reasoning which led to Eq.44, it can be easily shown that,

$$\sum_{j=1}^{m_{(s)}} w_{1,j}^{(t),(s)} = \sum_{j=1}^{m_{(s)}} w_{2,j}^{(t),(s)} = \ldots = \sum_{j=1}^{m_{(s)}} w_{m_{(t)},j}^{(t),(s)} = \psi_{t,s} \qquad \text{(C4)}$$

As in the case of the square Potts lattice, the maximum eigenvalue can be calculated from the reduced transfer matrix $\boldsymbol{\Psi}$ with the elements $\psi_{i,j}$.

Table 1. Comparison of the size of the reduced transfer matrix $\boldsymbol{\alpha}_r$ with that of the original transfer matrix $\boldsymbol{B}_r$ for the various limited triangular and square Ising lattices.

| The size of lattice (*r*) | Order of the Original transfer matrix | Order of the reduced transfer matrix |
|---|---|---|
| 3 | 8 | 4 |
| 4 | 16 | 6 |
| 5 | 32 | 8 |
| 6 | 64 | 13 |
| 7 | 128 | 18 |
| 8 | 256 | 29 |
| 9 | 512 | 46 |
| 10 | 1024 | 75 |
| 11 | 2048 | 126 |
| 12 | 4096 | 201 |
| 13 | 8192 | 374 |
| 14 | 16384 | 615 |
| 15 | 32768 | 1144 |
| 16 | 65536 | 1876 |

Table 2. Computed free energy per site, $(1/r)\ln\lambda_{\max}$, at the critical point for the various limited triangular and square Ising lattices, by using both the reduced transfer matrix $\boldsymbol{\alpha}_r$ and the original transfer matrix $\boldsymbol{B_r}$.

| | Square | | Triangular | |
|---|---|---|---|---|
| Lattice size ($r$) | Reduced transfer matrix | Original transfer matrix | Reduced transfer matrix | Original transfer matrix |
| 3 | 0.9611726693262 | 0.9611726693262 | 0.9049142122605 | 0.9049142122605 |
| 4 | 0.9467841410239 | 0.9467841410239 | 0.8937798187581 | 0.8937798187581 |
| 5 | 0.9404454417199 | 0.9404454417199 | 0.8886604380857 | 0.8886604380857 |
| 6 | 0.9370953533857 | 0.9370953533857 | 0.8858852815798 | 0.8858852815798 |
| 7 | 0.9351051975940 | 0.9351051975940 | 0.8842132030119 | 0.8842132030119 |
| 8 | 0.9338245734102 | 0.9338245734102 | 0.8831283104747 | 0.8831283104747 |
| 9 | 0.9329512955251 | 0.9329512955251 | 0.8823846302477 | 0.8823846302477 |
| 10 | 0.9323288953175 | 0.9323288953175 | 0.8818527261286 | 0.8818527261286 |
| 11 | 0.9318695628671 | 0.9318695628671 | 0.8814591970312 | 0.8814591970312 |
| 12 | 0.9315207079118 | 0.9315207079118 | 0.8811594483165 | 0.8811594483165 |
| 13 | 0.9312498843820 | 0.9312498843820 | 0.8809269979851 | 0.8809269979851 |
| 14 | 0.9310350360548 | 0.9310350360548 | 0.8807419695026 | 0.8807419695026 |
| 15 | 0.9308619702215 | 0.9308619702215 | 0.8805930569194 | 0.8805930569194 |
| 16 | 0.9307202659175 | 0.9307202659175 | 0.8804708453356 | 0.8804708453356 |

Table 3. Comparison of the size of the reduced transfer matrix $\boldsymbol{\Gamma}_r$ with that of the original transfer matrix $\boldsymbol{C}_r$ for various limited triangular and square three-state Potts models.

| The size of lattice ($r$) | Order of the transfer matrix | Order of the reduced transfer matrix |
|---|---|---|
| 3 | 27 | 6 |
| 4 | 81 | 13 |
| 5 | 243 | 21 |
| 6 | 729 | 49 |
| 7 | 2187 | 97 |
| 8 | 6561 | 230 |
| 9 | 19683 | 519 |
| 10 | 59049 | 1311 |
| 11 | 177147 | 3417 |

Table 4. $(1/r)\ln\lambda_{max}$ at the critical point for the various finite-size square Potts lattices, using both the original and reduced transfer matrices, compared with the result of de Queiroz[b].

| Lattice size (*r*) | Original transfer matrix | Reduced transfer matrix | Computed by others[b] |
|---|---|---|---|
| 3 | 2.1210912619804 | 2.1210912619804 | 2.121091261980 |
| 4 | 2.0977045200304 | 2.0977045200304 | 2.097704520030 |
| 5 | 2.0874606638059 | 2.0874606638059 | 2.087460663806 |
| 6 | 2.0820636899026 | 2.0820636899026 | 2.082063689903 |
| 7 | 2.0788633563350 | 2.0788633563350 | 2.078863356335 |
| 8 | 2.0768091780191 | 2.0768091780191 | 2.076806333203 |
| 9 | 2.0754081515407 | 2.0754081515407 | 2.075404683690 |
| 10 | 2.0744145949876 | 2.0744145949876 | 2.074406246134 |

[b]S. L. A. de Queiroz, *J. Phys. A.* **33**, 721 (2000).

Table 5. $(1/r)\ln\lambda_{max}$ at the critical point for the various finite-size triangular Potts lattices, using both the original and reduced transfer matrices, compared with the result of de Queiroz[b].

| Lattice size (*r*) | Original transfer matrix | Reduced transfer matrix | Computed by others[b] |
|---|---|---|---|
| 3 | 2.0029599009392 | 2.0029599009392 | 2.002959900939 |
| 4 | 1.9850089217698 | 1.9850089217698 | 1.985008921770 |
| 5 | 1.9767773674830 | 1.9767773674830 | 1.976777367483 |
| 6 | 1.9723212910682 | 1.9723212910682 | 1.972321291068 |
| 7 | 1.9696386612195 | 1.9696386612195 | 1.969638661220 |
| 8 | 1.9679059276432 | 1.9679059276432 | 1.967899052250 |
| 9 | 1.9667158163274 | 1.9667158163274 | 1.966707034799 |
| 10 | 1.9658729866941 | 1.9658729866941 | 1.965854709111 |

[b]S. L. A. de Queiroz, *J. Phys. A.* **33**, 721 (2000).